\documentstyle[epsf]{ioplppt}    
\begin{document}


\title{Periodic Orbits, Breaktime and Localization}
\author{
Doron Cohen \\ 
Department of Physics of Complex Systems, \\ 
The Weizmann Institute of Science, Rehovot 76100, Israel.
}


\begin{abstract}

The main goal of the present paper is to convince that 
it is feasible to construct a `periodic orbit theory' 
of localization by extending the idea of classical action 
correlations.  This possibility had been questioned by many 
researchers in the field of `Quantum Chaos'.  
Starting from the semiclassical trace formula, we formulate 
a quantal-classical duality relation that connects the 
spectral properties of the quantal spectrum to the 
statistical properties of lengths of periodic orbits.
By identifying the classical correlation scale it is possible 
to extend the semiclassical theory of spectral statistics,  
in case of a complex systems, beyond the limitations that are 
implied by the diagonal approximation. We discuss the quantal 
dynamics of a particle in a disordered system. The various 
regimes are defined in terms of time-disorder `phase diagram'. 
As expected, the breaktime may be `disorder limited'
rather than `volume limited', leading to localization if 
it is shorter than the ergodic time. Qualitative agreement 
with scaling theory of localization in one to three dimensions 
is demonstrated. 

\end{abstract}


\section{Introduction}

Extending the semiclassical approach to spectral statistics beyond
the diagonal approximation is presently one of the most vigorously 
pursued direction of research in ``quantum chaos''. It is desirable 
to reach a semiclassical understanding of the long-time behavior 
also for disordered systems. They play a central role in condensed-matter 
as well as in mesoscopic physics.  The introduction of semiclassical 
methods in the latter case is quite natural. It can be expected 
that a semiclassical insight into localization will, in turn, shed 
new light on semiclassical methods in general. The present paper is 
intended as a contribution towards this goal. It rests mainly on two 
previous observations: the connection between spectral correlations 
in the long-time regime and classical action-correlations \cite{argaman},   
and the heuristic treatment of localization by Allen \cite{allen}.
It turns out that the latter appears as a natural consequence of the 
former, once a disorder system is considered.  An improved qualitative 
picture of spectral statistics follows, expressed in the form of 
`time-disorder' diagram. Furthermore, the present formulation paves 
the way towards a quantitative account in terms of the spectral form 
factor. 
  
There are few time scales that are associated with the  
semiclassical approximation for the time-evolution 
of any observable. Such an approximation involves 
a double-summation over classical orbits. The purpose of 
the following paragraphs is to make a clear 
distinction between these various time scales. In particular
we wish to clarify the `breaktime' concept that plays 
a central role in our formulation. 
Initially the classical behavior is followed. One relevant 
time scale for the departure from the classical behavior 
is $t_{scl}$. By definition, when $t_{scl}<t$ deviations 
that are associated with the breakdown of the 
{\em stationary phase approximation} may show up. 
It has been argued \cite{heller} that $t_{scl}\sim \hbar^{-1/3}$. 
Further deviations from the leading order 
semiclassical expansion due to diffraction effects
are discussed in \cite{diffraction}. 
One should be careful not to confuse these deviations,  
which are associated with the accuracy of the stationary phase
approximation with the following discussion of the breaktime
concept. It is assumed in the sequel that the leading order 
semiclassical formalism constitutes a qualitatively good 
approximation also for $t_{scl}{<}t$ in spite of these deviations.  

{\em Interference effects} lead to further deviations from the 
classical behavior. Well isolated classical paths, for 
which the stationary phase approximation is completely accurate,   
still may give rise to either constructive or destructive 
interference effect. From now on we shall put the focus  
on the semiclassical computation of the spectral form factor 
\cite{berry}, where the double-summation is over classical 
{\em periodic orbits} (POs). We shall disregard extremely 
short times, for which only few POs contribute, since for 
any generic chaotic system the POs proliferate exponentially 
with time. The simplest assumption would be that the interference 
contribution (off-diagonal terms) is self-averaged to zero.  
However, such an assumption would imply that the classical 
behavior is followed for arbitrarily long times. 
This is obviously not true. After a sufficiently long time 
the discrete nature of the energy spectrum becomes apparent, 
and the {\em recurrent} quasiperiodic nature of the dynamics 
is revealed. The breaktime $t^*$ is the time scale which is 
associated with the latter crossover. Neglecting the 
interference contribution for $t<t^*$ is known as the  
diagonal-approximation \cite{berry}.

From semiclassical point of view the breaktime $t^*$ is 
related to the breakdown of the {\em diagonal approximation}.
It has been conjectured that the breakdown of the 
diagonal approximation is a manifestation  
of classical action-correlations \cite{argaman}. 
Else, if the actions were uncorrelated (Poisson statistics),  
then the off-diagonal (interference) contribution would be 
self-averaged to zero. Typically the breaktime $t^*$ is 
identified with the Heisenberg time $t_H=2\pi\hbar/\Delta E$, 
where $\Delta E$ is the average level spacing.   
The Heisenberg time is semiclassically 
much longer than the `log' time $t_E \sim \ln(1/\hbar)$ over  
which classical orbits proliferates on the 
uncertainty scale. The latter time scale has no 
physical significance as far as the form factor is concerned. 
See also discussion after Eq.(\ref{e_14}).

The breaktime which is determined by the Heisenberg uncertainty relation  
is volume dependent. However, for a disordered system the breaktime 
may be much shorter and volume-independent due to localization effect.
The theory for this this `disordered limited' rather than 
`volume limited' breaktime constitutes the main theme of the 
present paper. 
Our approach to deal with disorder within the framework 
of the semiclassical approach constitutes a natural extension 
of previous attempts to integrate `mesoscopic physics' with 
the so called field of `quantum chaos'. See \cite{thomas}
for review.  Note that a naive semiclassical arguments can be 
used in order to estimate the breaktime and the localization length 
for 1D systems \cite{shep}. See the next section for further details. 
This argument, as it stands, cannot be extended to higher dimensions, 
which implies that a fundamentally different approach is needed.  
The same objection applies to a recent attempt to 
propose a periodic orbit theory for 1D localization \cite{scharf}.
In the latter reference the semiclassical argument for localization 
is based on proving exponentially-small sensitivity for change in boundary 
conditions.  This is due to the fact that only exponentially 
small number of POs with $t{<}t_H$ hit the edges. The statement 
holds for $d{=}1$, where $t_H{\ll}t_{erg}$, but it fails in higher 
dimensions. Hence the necessary condition for having localization 
should be much weaker.

The plan of this paper is as follows. The expected results for 
the disordered limited breaktime, based on scaling theory of localization, 
are presented in Section 2.  Our main goal is to re-derive these results 
from semiclassical consideration. In section 3, starting from the 
semiclassical trace formula (SCTF), we formulate a duality relation  
that connects the spectral properties of the quantal spectrum to 
the classical two-point statistics of the POs.  In section 4 we 
identify the classical correlation scale. Then it is possible 
to extend the semiclassical theory of spectral statistics,  
in case of a complex systems, beyond the limitations that are 
implied by the diagonal approximation. 
In section 5 we demonstrate that a disorder limited 
breaktime is indeed a natural consequence of our formulation.  
The various time regimes for a particle in a disordered system 
are illustrated using a  time-disorder `phase diagram'. 
Localization show-up if there is a disorder limited breaktime    
which is shorter than the ergodic time. Semiclassical 
interpretation for the existence of a critical and an ohmic regimes 
for 3D localization is also introduced. Finally, in section 6, we 
introduce a semiclassical approximation scheme for the 
form factor, that goes beyond the diagonal approximation. 
The limitations of this new scheme are pointed out.

Effects that are associated with the actual presence of a magnetic 
field are not considered in this paper, since the SCTF should be 
modified then. Still, for simplicity of presentation we 
cite for the form factor the GUE rather than the GOE result, 
and we disregard the effect of time reversal symmetry.
A proper treatment of these details is quite obvious, 
and will appear in a future publication \cite{spc}. 
It is avoided here in order not to obscure the main point.

\section{Breaktime for Disordered Systems}   
  
We consider a particle in a disordered potential. 
The classical dynamics is assumed to be diffusive. 
For concreteness we refer to a {\em disordered billiard}.
The concept is defined below. It should be emphasized 
that we assume genuine disorder. Pseudo-random 
disorder, as well as spatial symmetries are out of the scope 
of our considerations.   

A disordered billiard is a quasi $d$-dimensional 
structure that consists of connected chaotic cavities. 
Here we summarize the parameters that are associated with 
its definition. The billiard is embedded in $d_0$-dimensional 
space ($2\le d_0$). It constitutes a $d$ dimensional 
structure of cells, (obviously $d\le d_0$). 
Each cell, by itself, constitutes a chaotic cavity whose 
volume is roughly $\ell_0^{d_0}$. 
However, the cells are connected by small holes whose 
area is $a_0^{d_0{-}1}$ with $a_0\ll\ell_0$.
The volume of the whole structure is $\Omega={\cal L}^d\ell_0^{d_0{-}d}$.
Assuming a classical particle whose velocity is $v$,    
the average escape time out of a cell is 
$t_0 \approx (\ell_0/v)\cdot(\ell_0/a_0)^{d_0{-}1}$.
The classical diffusion coefficient is ${\cal D}_0 = \ell_0^2/t_0$. 
The classical diffusion law is 
$\langle(x-x_0)^2\rangle = {\cal D}_0 t$ where $x_0$ is 
the location of an initial distribution.  

The mass of the particle is $m$. 
Its De-Broglie wavelength $\lambda_B=\hbar/mv$ 
is assumed to be much shorter than $\ell_0$ as to allow 
(later) semiclassical considerations. Actually, in order to 
have non-trivial dynamical behavior $\lambda_B$ should be smaller 
or at most equal to $a_0$ (note the following definition of 
the dimensionless conductance).      
The mean energy level spacing is 
$\Delta E=2\pi\alpha^{-1}{\cdot}(\hbar^{d_0}/{\cal L}^d)$ where 
$\alpha\sim\ell_0^{d_0{-}d}m^{d_0{-}1}v^{d_0{-}2}$.
The Heisenberg time is 
$t_H=2\pi\hbar/\Delta E = \alpha {\cal L}^d/\hbar^{d_0{-}1}$
The dimensionless conductance $g_0$ on scale of {\em one} cell
is defined as the ratio of the Thouless energy 
$2\pi\hbar/t_0$ to the level spacing $2\pi\hbar/t_H$. 
(for $t_H$ one should substitute here ${\cal L}=\ell_0$).
Hence $g_0=(a_0/\lambda_B)^{d_0{-}1}$ is simply 
related to the hole size.
Out of the eight independent parameters
$(d_0, d, \ell_0, a_0, {\cal L}, m, v, \hbar)$ there 
are actually only three dimensionless parameters 
which are relevant. Setting $t_0$ and $\ell_0$ to unity,
these are $d$, $g_0$ and ${\cal L}$. All the results should 
be expressed using these parameters.

For a billiard system in $d$-dimension, whose volume is $\Omega$, 
the Heisenberg-time is given by the expression 
$t_H = \alpha\Omega/\hbar^{d{-}1}$.
For the {\em disordered billiard} Heisenberg time can be expressed 
in terms of the unit-cell dimensionless conductance
\begin{eqnarray} \label{e1}
t^*_H \ = \ \frac{2\pi\hbar}{\Delta E} \ = \ 
\alpha\frac{\Omega}{\hbar^{d_0{-}1}} \ = \ 
\left(\frac{{\cal L}}{\ell_0}\right)^d g_0t_0
\end{eqnarray}
The actual `disordered limited' breaktime may be much shorter 
due to localization effect. Naive reasoning concerning 
wavepacket dynamics leads to the 
volume-independent estimate \cite{shep} 
\begin{eqnarray}\label{e2}
t^* \ = \ \frac{2\pi\hbar}{\Delta_\xi} 
\ = \ \alpha \frac{\xi^d}{\hbar^{d_0{-}1}} 
\ \ \ \ \ [\mbox{naive}] \ .
\end{eqnarray}
Here $\Delta_\xi$ is the effective level spacing within 
a volume $\xi^d$. Assuming that up to $t^*$ the spreading is
diffusive-like, it follows that 
\begin{eqnarray} \label{e3}
\xi^2 \ = \ {\cal D}_0 \  t^*     
\end{eqnarray}
Combining these two equations it has been argued \cite{shep} that 
for quasi one-dimensional structure ($d{=}1$) the localization 
length is $\xi{\sim}\alpha{\cal D}_0/\hbar^{d_0{-}1}$. 
In terms of the dimensionless conductance the result is 
$\xi{=}g_0\ell_0$ which corresponds to the breaktime 
\begin{eqnarray} \label{e4}
t^* \ = \ g_0^2 \ t_0  
\ \ \ \ \ [\mbox{for} \ \ d{=}1] \ .
\end{eqnarray} 
The above argument that relates $\xi$ and $t^*$ to 
the dimensionless conductance $g_0$ cannot 
be extended in case of $1{<}d$. This is due to 
the fact that (\ref{e2}) overestimates the breaktime.
From scaling theory of localization \cite{scaling} one 
obtains for $d{=}2$ the result $\xi{=}\mbox{e}^{g_0}\ell_0$
leading (via equation (3)) to 
\begin{eqnarray} \label{e5}  
t^* \ = \ \mbox{e}^{2g_0} \ t_0  
\ \ \ \ \ [\mbox{for} \ \ d{=}2] \ .
\end{eqnarray}
For $d{=}3$ and $g{<}g_c$, where $g_c$ is the critical value
of $g$, scaling theory predicts $\xi{=}|g{-}g_c|^{-\nu}\ell_0$,
with $\nu{\approx}1/(d{-}2)$.
Here the diffusive-like behavior up to $t^*$ is replaced by 
an anomalous scale-dependent diffusive behavior, 
leading to the relation $\xi^d{=}\ell_0^{d{-}2}{\cal D}_0t^*$
rather than (\ref{e3}), and hence 
\begin{eqnarray} \label{e6}
t^* \ = \ \frac{1}{|g_0-g_c|^{\nu d}} \ t_0 
\ \ \ \ \ [\mbox{for} \ \ d{=}3] \ .
\end{eqnarray} 
It is easily verified that the naive formula (\ref{e2}) overestimates 
the actual breaktime by factor $g_0$ for both $d{=}2$ and $d{=}3$.
We turn now to develop a semiclassical theory for the breaktime.

\section{Quantal-Classical Duality}

The Semiclassical trace formula (SCTF) \cite{trace} relates the 
quantal density of states to the classical density of 
periodic orbits (POs).
The quantal spectrum $\{k_n\}$  for a simple billiard in $d$ dimensions
is defined by the Helmholtz equation $(\nabla^2{+}k^2)\psi=0$ 
with the appropriate boundary conditions. 
The corresponding quantal density is
\begin{equation}
\rho_{qm}(k) \equiv 
\sum_n 2\pi\delta(k{-}k_n) \ |_{osc} \ \ \ \ .
\end{equation}
In order to facilitate the application of Fourier 
transform conventions a factor $2\pi$ has been incorporated
and let $\rho(k){=}\rho(-k)$ for $k{<}0$. The subscript $osc$ 
implies that the averaged (smoothed) density of states
is subtracted. This smooth component equals the 
corresponding Heisenberg length and is found 
via Weyl law, namely
\begin{eqnarray}   \label{e_8}
L_H(k) \ = \ C_d \Omega \ k^{d{-}1} 
\end{eqnarray}
where $\Omega$ is the volume of the billiard,
and $C_d=(2^{d{-}2}\pi^{{d}/{2}{-}1}\Gamma({d}/{2}))^{-1}$.  
For billiard systems, actions lengths and times are trivially 
related by constant factors and therefore can be used 
interchangeably. In the sequel some of the formulas become 
more intelligible if one recalls that $L$ plays actually the 
role of the time.
The classical spectrum $\{L_j\}$ consists of the lengths 
of the POs and their repetitions. 
The corresponding weighted density is 
\begin{equation}
\rho_{cl}(L) \equiv
\sum_j A_j \delta(L{-}L_j) \ |_{osc} \ \ \ \ .
\end{equation}
Here $A_j$ are the instability amplitudes.
We note that for a simple chaotic billiard, due 
to ergodicity  
$K_D(L)\equiv \langle 
\sum_j |A_j|^2\delta(L{-}L_j) 
\rangle \sim L$. 
The instability amplitudes decay exponentially
with $L$, namely $|A_j|^2{\sim}L^2\exp(-\sigma L)$, where
$\sigma$ is the Lyapunov exponent. Hence the density 
of POs grows exponentially 
as $\exp(\sigma L)/L$. 
with the above definitions the SCTF is simply
\begin{equation}
\rho_{qm}(k) \ \ = \ \ {\cal FT} \ \rho_{cl}(L) \ \ \ \ .
\end{equation}
Where the notation ${\cal FT}$ is used in order 
to denote a Fourier transform. Both the SCTF and 
the statistical relation (\ref{e_ff}) that follows, 
reflect the idea that the quantal spectrum and the 
classical spectrum are two dual manifestations 
of the billiard boundary.

The two-points correlation function of the quantal spectrum is 
$R_{qm}(k,\epsilon)\equiv
\langle\rho_{qm}(k)\rho_{qm}(k{+}\epsilon)\rangle$,
where the angle brackets denote statistical averaging.
The spectral form factor $K_{qm}(k,L)$ is its Fourier 
transform in the variable $\epsilon{\leadsto}L$.
Due to the self correlations of the discrete energy 
spectrum $R_{qm}(\epsilon)$ is delta-peaked 
in its origin. As a consequence the asymptotic behavior 
of the spectral form factor is $K_{qm}(k,L)=L_H(k)$
for $L_H(k){\ll}L$. For a {\em simple} ballistic billiard the 
crossover to the asymptotic behavior occurs indeed  
at the Heisenberg time. The functional form of the 
crossover is described by Random Matrix Theory (RMT). 
For concreteness we cite the approximation 
$K_{qm}(k,L){=}\min(L,L_H(k))$. (Effect of symmetries 
is being ignored for sake of simplicity). 
In order to formulate a semiclassical theory for the 
form factor it is useful to define the two point 
correlation function of the classical spectrum
$R_{cl} (x,L) \equiv  
\langle \rho_{cl}(L) \rho_{cl}(L{+}x) \rangle$. 
The corresponding form factor $K_{cl}(k,L)$ is obtained 
by Fourier transform in the variable $x{\leadsto}k$.
It is straightforward to prove that due to  
the SCTF $R_{qm}(k,\epsilon)$ is related to 
$R_{cl}(x,L)$ by a double Fourier transform. Hence 
\begin{eqnarray} \label{e_ff}
K_{qm}(k,L) \ = \ K_{cl}(k,L) \ \ ,
\end{eqnarray}  
which is the two point version of the SCTF. It constitutes
a concise semiclassical relation that expresses 
the statistical implication of quantal-classical duality.    
It is essential to keep the spectral form factor 
un-rescaled. Its parametric dependence should not be 
suppressed. If regarded as a function of $L$, the quantity 
$K(k,L)$ is the quantal form-factor, while if regarded as 
a function of $k$ it is the classical form factor.

\section{Beyond the Diagonal Approximation}

The two points statistics of the quantal density 
reflects the discrete nature of the quantal spectrum, 
and also its rigidity.  
It follows that the classical spectrum should 
be characterized by non-trivial correlations 
that can be actually deduced from (\ref{e_ff}). 
This type of argumentation has been used in \cite{argaman} 
and will be further developed here. It is useful to write 
$R_{cl}(x,L)=K_D(L)(\delta(x){-}p(x))$ where a non-vanishing
$p(x)$ implies that the classical spectrum is characterized by 
non-trivial correlations. Note that a proper treatment 
of time reversal symmetry is avoided here.  
Denoting the classical correlation 
scale by $\lambda(L)$ it follows that $K(k,L)$ should 
have a breaktime that is determined 
via $k\sim{2\pi}/\lambda(L)$. For a {\em simple} ballistic 
billiard this should be equivalent to $L\sim L_H(k)$. 
Thus we deduce that the classical correlation scale is
\begin{eqnarray} \label{e_14}
\lambda(L)=2\pi\left(C_d\frac{\Omega}{L}\right)
^{\frac{1}{d{-}1}} \ \ \ .
\end{eqnarray}  
If ${2\pi}/{\lambda} \ll k$ then $K(k,L) \approx K_D(L)$, 
which is the diagonal approximation. More generally
$K(k,L) = C(k,L)K_D(L)$,
where $C(k,L) = (1{-}\tilde{p}(k))$ 
and $\tilde{}$ denotes Fourier transformed density. 
Note that it is implicit that both $p(x)$ and $\tilde{p}(k)$ 
depend parametrically on $L$.    
For $L_H(k) \ll L$ one should 
obtain the correct asymptotic behavior $K(k,L) = L_H(k)$. 
Therefore $C(k,L) \rightarrow 0$ in this regime and 
consequently the normalization $\int_{-\infty}^{+\infty}p(x)dx=1$ 
should be satisfied. 
It is natural to introduce a scaling function such that 
$p(x) = \lambda^{-1}\hat{p}(x/\lambda)$ and consequently
$C(k,L) = \hat{C}(k\lambda(L))$. For a simple ballistic billiard, 
neglecting modifications due to time reversal symmetry,   
the scaling function 
$\hat{C}_{ballistic}(\kappa) = \min((\kappa/2\pi)^{d{-}1},1)$ 
will generate the correct quantum mechanical result. 
The related scaling function $\hat{p}(s)$ can be deduced 
via inverse Fourier transform of $(1{-}\hat{C}(\kappa))$.

It should be clear that the actual quantum-mechanical breaktime 
is related to the breakdown of the diagonal approximation. This 
breaktime is determined by the condition $k\lambda(L){\sim}2\pi$. 
If one confused $\lambda(L)$ with the classical spacing 
$\Delta L{\sim}L\exp(-\sigma L)$, then one would deduce 
a false breaktime at the `log' time $t_E\sim\ln(k)$.

A heuristic interpretation of the classical two points
statistics is in order. The normalization of $p(x)$ 
implies rigidity of the classical spectrum on large 
scales. Expression (\ref{e_14}) for the correlation scale 
is definitely not obvious. 
Still, the length scale $\lambda$ possess a very simple 
geometrical meaning.  It is simply the typical distance 
between neighboring points where the PO 
had hit the billiard surface.
It is important to notice that $\lambda$ is much 
larger than the average spacing of the classical 
spectrum. The latter is exponentially small in $L$
due to the exponential proliferation of POs.
This fact suggest that the overwhelming majority 
of POs is uncorrelated with a given reference PO. 
The POs that effectively contribute to $p(x)$ must be 
geometrically related in some intimate way.
Further discussion of these heuristic observations 
will be published elsewhere \cite{spc}.

For a billiard that is characterized by a complicated structure, 
the ergodic time is much larger than the ballistic time. 
Orbit whose period $L_j$ is less than the 
ergodic time will not explore the whole volume of the 
billiard but rather a partial volume $\Omega_j$. It is
quite obvious that POs that does not explore the {\em same}
partial volume cannot be correlated in length, unless
some special symmetry exists. The possibility to make 
a classification of POs into statistically independent 
classes constitutes a key observation for constructing 
an approximation scheme that goes beyond the diagonal
approximation.  
Due to the classification, the spectral 
form factor can be written as a sum 
$K(k,L)=\sum_{\Omega}K(k,L,\Omega)$ of statistically 
independent contributions, where $K(k,L,\Omega)$ involves 
summation over POs with $\Omega_j{\sim}\Omega$. 
Thus the following semiclassical expression is obtained
\begin{eqnarray} \label{e_ff2}
K(k,L) \ = \ \sum_j \ \hat{C}(k\lambda_j)|A_j|^2\delta(L{-}L_j)
\end{eqnarray} 
with $\lambda_j$ that corresponds to the explored volume $\Omega_j$ 
and with scaling function $\hat{C}(\kappa)$ that may depend on 
the nature of the dynamics. 
This formula constitutes the basis for our theory.

\section{Theory of Disordered Billiards} 

We turn now to apply semiclassical considerations 
concerning the dynamics of a particle in 
a disordered system. The classical dynamics is assumed 
to be diffusive and we refer again to the 
{\em disordered billiard} of section 2. It should be 
re-emphasized that we assume genuine disorder. 
Pseudo-random disorder, as well as spatial symmetries 
may require a more sophisticated theory of PO-correlations.

From now on we translate lengths into times by using $L{=}vt$. 
The diagonal sum over the POs satisfies $K_D(t)= tP_{cl}(t)$,
where $P_{cl}(t)$ is the classical `probability' to return
\cite{tau}. For ballistic billiard  $P_{cl}(t)=1$. This is 
true also for diffusive systems provided $t_{erg}{<}t$, where 
\begin{eqnarray} \label{e16}
t_{erg} \ = \ \frac{{\cal L}^2}{{\cal D}_0} 
\ = \ \left(\frac{{\cal L}}{\ell_0}\right)^2 t_0  
\end{eqnarray}
For $t < t_{erg}$ the classical probability to return is  
$P_{c\ell}(t) = (t_{erg}/(2\pi t))^{d/2}$.  The latter 
functional form reflects the diffusive nature of the 
dynamics.

The POs of a the disordered billiard can be classified 
by the volume $\Omega_j$ which they explore. By definition 
$\Omega_j$ is the total volume of those cells that were 
visited by the orbit. Let us consider POs whose 
length is $t$. Their probability distribution with respect 
to the explored volume will be denoted by $f_t(\Omega)$. 
This distribution can be deduced from purely classical 
considerations. The detailed computation for the special 
case of $d{=}1$ will be published elsewhere \cite{spc}. 
The result is, 
\begin{eqnarray} \label{e_fx}
f_t(x) \ = \ \frac{1}{\sqrt{t/t_0}\Omega_0} \ 
\hat{f}\left(\frac{\Omega}{\sqrt{t/t_0}\Omega_0}\right) 
\ \ \ \ \ \ \mbox{[for $d{=}1$]} ,
\end{eqnarray}
where $\Omega_0 = \ell_0^{d_0}$ is the unit-cell volume.
The scaling function $\hat{f}(x)$ satisfies 
$\int_0^{\infty}\hat{f}(x)dx=1$ and
$\int_0^{\infty}x\hat{f}(x)dx=\sqrt{\pi/2}$. 
It is plotted in Figure 1.

\begin{figure} 
\begin{center}
\leavevmode 
\epsfysize=2.0in
\epsffile{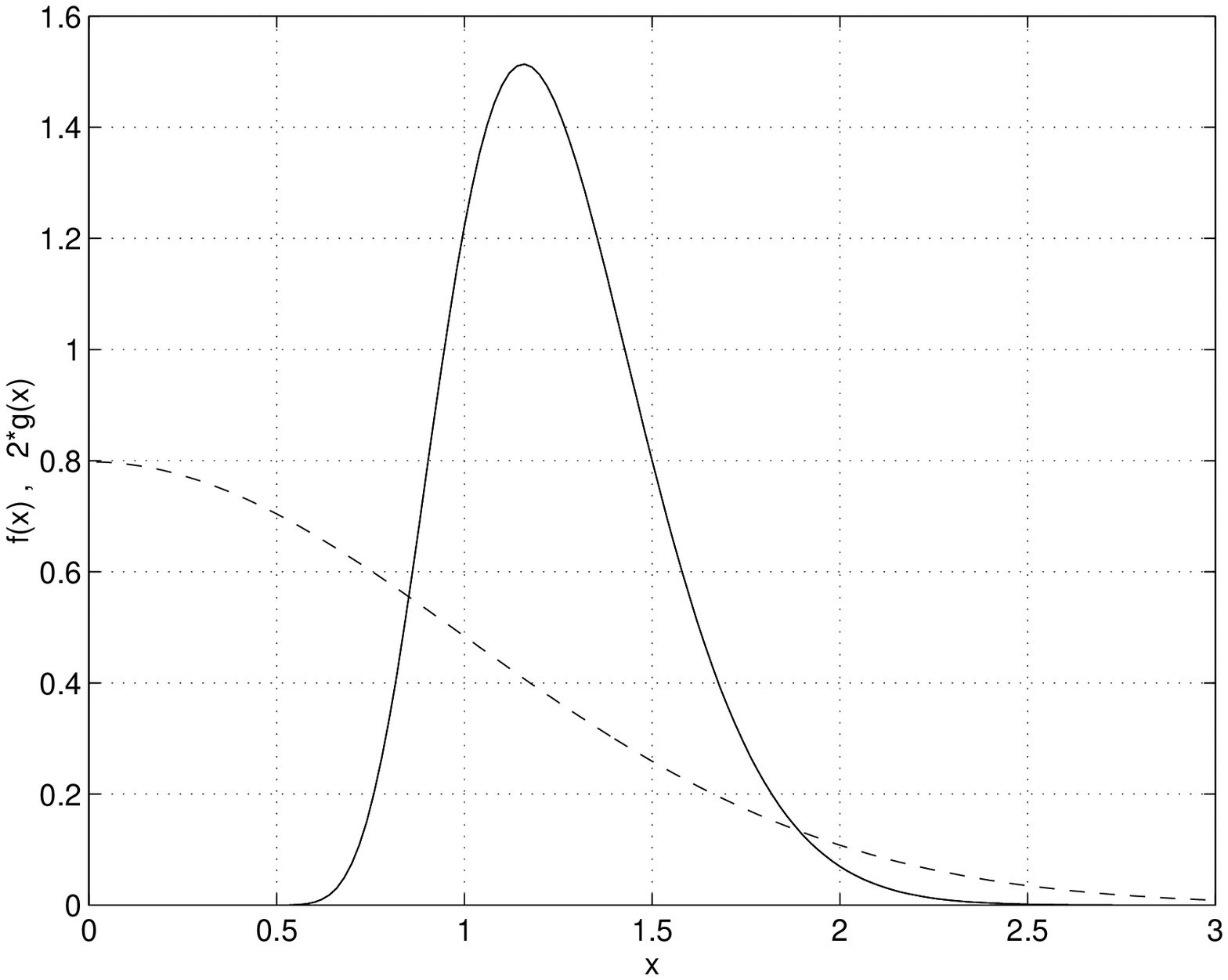}
\epsfysize=2.0in
\epsffile{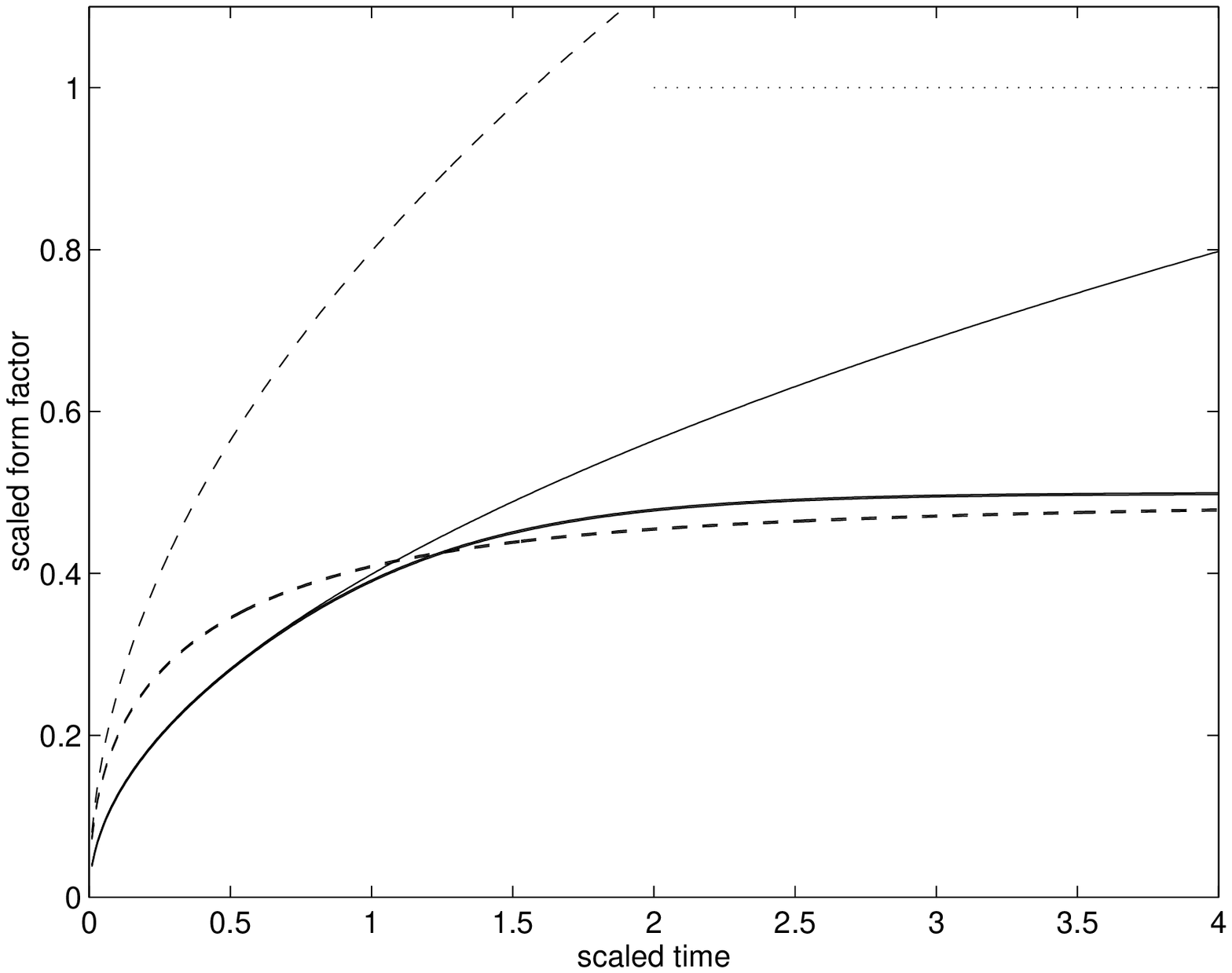}
\end{center}
\caption{\protect\footnotesize 
Left plot: The scaled probability distribution of 
the explored volume for a disordered chain (solid line), 
compared with a Gaussian distribution that characterizes 
the diffusion profile (dashed line). 
Right plot: The scaled form factor for a disordered 
infinite chain. Solid line - GUE result,
Dashed line - GOE result. 
The thiner curves are obtained by employing 
the diagonal approximation, while the thicker
curves are obtained by employing the BLC 
approximation scheme. The dotted line 
illustrates the correct asymptotic behavior. } 
\label{f_explore}
\end{figure}

The average volume which is explored by a POs of 
length $t$ will be denoted by $\Omega_e(t)$.
For $d{=}1$ obviously $\Omega_e(t) \propto \sqrt{t}$,
while for $d{>}2$ the average volume which is explored after 
time $t$ is $\Omega_e(t) \sim t$ to leading order. 
Specifically, one may write 
$\Omega_e(t) = \ell_0^{d_0}(t/t_0)F(t/t_0)$
where, following \cite{weiss},
\begin{eqnarray} \label{e18}
F(\tau)=\left\{ \matrix{
\sqrt{\frac{8}{\pi}}\frac{1}{\tau^{1/2}} & \mbox{for} \ \  d{=}1  \cr
\pi \frac{1}{\ln(\tau)} & \mbox{for} \ \  d{=}2  \cr
c+\frac{c'}{\tau^{(d{-}2)/2}} & \mbox{for} \ \  2{<}d{<}4  \cr
c+\frac{c''}{\tau} & \mbox{for} \ \  4{<}d   } \right. 
\end{eqnarray}   
Above $c$ and $c'$ and $c''$ are constants 
of order unity (for simple cubic-like structure $c\sim 0.7$). 
Note that the numerical prefactor $\sqrt{8/\pi}$ 
for the $d=1$ case in (16) is somewhat larger than 
the $\sqrt{\pi/2}$ which is implied by (15). This 
difference is probably due to the fact that (16) is 
not an exact result if POs are concerned, rather it 
is an exact result for wandering trajectories.
The transient time $t_0$ is actually a statistical entity, 
hence, associated with $\Omega_e(t)$ one should consider 
a dispersion 
$\Delta \Omega_e(t)\approx\ell_0^{d_0}{\cdot}\sqrt{t/t_0}$.
Note that in case of Eq.(\ref{e_fx}), the average 
explored volume and its dispersion are derived from 
a one-parameter scaling relation. This is not the case 
for $1<d$ diffusive system.

It is essential to distinguish the average explored 
volume $\Omega_e(t)$ from the diffusion volume $\Omega_d(t)$. 
The latter is determined by the diffusion law 
$\Omega_d(t)=\ell_0^{d_0{-}d}({\cal D}_0t)^{d/2}$.
The diffusion volume $\Omega_d$ refers to the instantaneous 
profile of an evolving distribution. It equals roughly 
to the total volume of those cells which are occupied by 
the evolving distribution.   
Note that $P_{cl}(t) \approx \Omega/\Omega_d(t)$.

Given the distribution $f_t(\Omega)$, expression 
(\ref{e_ff2}) can be cast back into the concise form 
$K(k,t)=C(k,t)K_D(t)$ with 
\begin{eqnarray} \label{e_corr}
C(k,t) \ = \ \int_{0}^{\infty}f_t(\Omega)d\Omega
\ \hat{C}(k\lambda(\Omega,t)) \ \ .
\end{eqnarray} 
The diffusive behavior that corresponds to the 
diagonal approximation prevails as long as 
the  condition $2\pi < k\lambda(\Omega_e(t))$ 
is satisfied. This condition can be cast into 
the more suggestive form  $t < t_H(\Omega_e(t))$. 
The equivalence of latter inequality with the former 
should be obvious from the discussion of the 
classical correlation scale in section 4.
(There we had taken the reverse route in order 
to deduce the expression for the classical 
correlation scale that correspond to a simple 
ballistic billiard). 
The concept of {\em running Heisenberg time} 
$t_H(\Omega_e(t))$ emerges in a natural way 
from our semiclassical considerations. 
Originally, this concept has been introduced
on the basis of a heuristic guess by Allen. 
In his paper \cite{allen} the concept appears in 
connection with the tight 
binding Anderson model where the on site energies are 
distributed within range $V$ and the hoping probability 
is $W$. There $g_0 \sim W/V$. Allen has pointed out that 
a qualitative agreement with the predictions of scaling 
theory is recovered if $\xi^d$ in (\ref{e2}) is replaced 
by $\Omega_e(t)$ as in our formulation. 
The condition $t<t_H(\Omega_e(t))$ for having a 
diffusive-like behavior can be cast into 
the form $F(t/t_0)>1/g_0$. It is easily verified 
that for both $d{=}1$ and $d{=}2$ the results for the breaktime 
are consistent with (\ref{e4}) and (\ref{e5}). For $d{=}3$ the 
existence of critical conductance  $g_c=c^{-1}$ is 
a natural consequence, but the exponent in (\ref{e6}) 
is $2/(d{-}2)$ rather than $\nu d$.


\begin{figure} 
\begin{center}
\leavevmode 
\epsfysize=4.0in
\epsffile{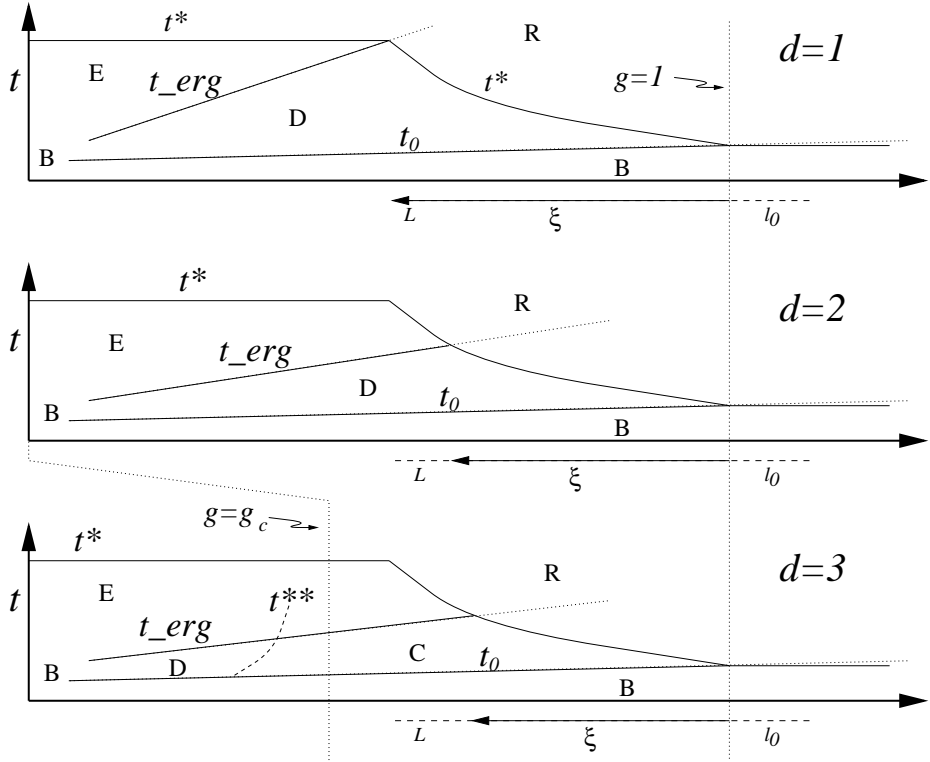}
\end{center}
\caption{\protect\footnotesize 
The different time 
regimes for  quantal evolution (vertical axis) 
versus disorder (horizontal axis), 
for one-two-three dimensional system. 
The ballistic (B), diffusive (D), critical (C), 
ergodic (E) and Recurrence (R) regimes are labeled. 
See further explanations of these diagrams 
in the text. }
\end{figure} 


Figure 2 illustrates the different time regimes  
for quantal evolution versus disorder for $d=1,2,3$. 
These diagrams constitute an improvement \cite{comp} over those of
\cite{aronov} and \cite{thomas}. 
For `zero disorder' $t_0$ may be interpreted as
the ballistic time scale that corresponds to the 
shortest PO. The breaktime is volume limited and determined 
by the Heisenberg time (\ref{e1}). As the disorder grows larger, 
two distinct classical time scale emerge, now $t_0$ is the ergodic 
time with respect to one cell, and $t_{erg}$ is the actual 
time for ergodicity over the whole volume. The latter 
is determined by the diffusion coefficient as in (\ref{e16}). If 
the disorder is not too large, the breaktime is still limited 
by the Heisenberg time.  Going to the other extreme 
limit of very large disorder ($g_0<1$) is not very interesting 
since the particle will be localized within the volume of 
a single cell.  For weaker disorder there is a crossover 
from a diffusive-like behavior (which is actually anomalous 
for $d{=}3$) to localization. The crossover time is determined 
by equations (\ref{e4}-\ref{e6}). In one dimension the crossover 
from `Heisenberg limited' breaktime to `disorder limited' 
breaktime happens to coincide 
with the classical curve for $t_{erg}$.  This 
coincidence does not occur in $d{=}2$ and therefore 
we have an intermediate  regime where the breaktime
occurs {\em after}  ergodization, but is still 
disorder limited rather than volume limited.
In three dimensions we have a qualitatively new 
regime $g_c<g_0$ where a purely diffusive (ohmic)
behavior (rather than diffusive-like behavior) prevails. 
The breaktime here is 
volume limited. Still, the border between the 
ohmic regime and the so-called `critical' one is 
non-trivial. Scaling theory predicts that the 
ohmic behavior is set only after a transient 
time $t^{**}$ which is given by (\ref{e6}) with 
$g_c{<}g_0$. In order to give a semiclassical 
explanation for $t^{**}$ we should refine somewhat
our argumentation.  The condition for {\em purely ohmic} 
behavior becomes $t<t_H(\Omega_e(t){-}\Delta\Omega_e(t))$. 
If $g_0$ is close to $g_c$ then there will appear 
a transient time $t<t^{**}$ where the {\em bare}
diagonal approximation is unsatisfactory. 
Note however that the critical exponent
turns to be by this argumentation $2$ rather 
than $\nu d$.

\section{The BLC approximation scheme}

We focus our attention on the actual computation 
of the form factor $K(\tau)$. 
Irrespective of any particular assumption
it is easily verified that $K(t_0){=}t_H/g_0$, 
while the asymptotic value $K_{qm}(t){=}t_H$ should 
be obtained for sufficiently long time. 
The asymptotic behavior reflects the discrete 
nature of the quantal spectrum. 
This feature imposes a major restriction on the 
functional form of $\hat{C}(\kappa)$.
Using $\hat{C}_{ballistic}(\kappa)$ (see discussion 
after (\ref{e_14})) one obtains that for a {\em simple} ballistic 
billiard there is a scaling function such that 
$K(t)=\Omega\hat{K}_0(t/\Omega)$, where $\Omega$ 
is the total volume. The correct 
asymptotic behavior is trivially obtained since 
by construction $\hat{C}_{ballistic}(\kappa)$ gives 
the correct quantum mechanical result.

For a disordered quasi-1D billiard, in order to determine 
the form factor, we should substitute (\ref{e_fx}) into 
(\ref{e_corr}). However, also the scaling function 
$\hat{C}(\kappa)$ should be specified. In order to make further 
progress towards a quantitative theory let us assume 
that it is simply equal to $\hat{C}_{ballistic}(\kappa)$. 
Using this assumption of ``ballistic like correlations'' (BLC) 
one obtains that the form factor satisfies the expected 
scaling property $K(t)=\Omega\hat{K}_d(t)$.
The latter scaling property, which implies the existence 
of a characteristic scaling function $\hat{K}_d(\tau)$,  
distinguishes a system with localization.  
Using the BLC approximation scheme the calculated 
$\hat{K}_1(\tau)$ is 
\begin{eqnarray}
\hat{K}_1(\tau) \ = \ \frac{1}{\sqrt{2\pi}} \int_0^{\infty} dx 
\ f\left(\frac{x}{\sqrt{\tau}}\right) \ \frac{x}{\tau} 
\hat{K}_0\left(\frac{\tau}{x}\right) 
 \ \ \ \ .
\end{eqnarray}   
It is plotted in Figure 1.  
Indeed the breaktime is disorder limited rather 
than volume limited. However, one observes that the 
computation yields the asymptotic behavior  
$K_{qm}(t){=}t_H/2$ rather than $K_{qm}(t){=}t_H$. 
This implies that the classical correlations have 
been overestimated by the BLC approximation scheme. 

The BLC approximation scheme can be applied for the 
analysis of $1<d$ localization. As in the $d{=}1$ case, 
the correct asymptotic behavior is {\em not} obtained.
The only way to guarantee a correct asymptotic 
behavior is to conjecture that 
$\hat{C}(\kappa) = (\Omega_d(t)/\Omega_e(t))\hat{C}_{ballistic}(\kappa)$
for $\kappa{<}1$. 
This required assumption implies that in spite of the 
net repulsion, the classical spectrum is further characterized 
by strong clustering. The effective clustering may be interpreted as 
arising from leaking of POs via `transverse' holes, thus
leaving out bundles of POs. 
Note that the normalization of $p(x)$
does not hold due to the leaking. Therefore 
$\hat{C}(\kappa)$ is modified in a way that is not 
completely compatible with its ballistic scaling form. 

For completeness we note that in the `critical regime' 
of $d=3$ localization, it has been 
suggested \cite{aronov} to put by hand the information
concerning the anomalous sub-diffusive behavior known 
from scaling theory. One obtains 
$\mbox{``}P_{c\ell}(t)\mbox{''}\sim({\cal L}/\ell_0)^d{\cdot}(t_0/t)$
and hence $K(t)$ is essentially the same as for $d{=}2$
system. The semiclassical justification for this procedure is 
not clear (see however Ref \cite{lerner}).
We believe that a better strategy would be to find 
the functional form of $f_t(\Omega)$ and $\hat{C}(\kappa)$ 
and to use~(\ref{e_corr}).

\section{Concluding Remarks}
 
We have demonstrated that simple semiclassical 
considerations are capable of giving an explanation 
for the existence of a disordered limited breaktime. 
Qualitatively, the results for the breaktime 
were in agreement with those of scaling theory of 
localization. In the last section we have briefly discussed  
the question whether future quantitative 
semiclassical theory for localization is feasible.
It turns out that the simplest (BLC) approximation 
scheme overestimates the rigidity of the 
classical spectrum.      

\ack{
I thank Harel Primack and Uzy Smilansky for interesting 
discussions. Thomas Dittrich and Shmuel Fishman  
are acknowledged for their comments. One of the referees 
is acknowledged for contributing the first paragraph of 
this paper. This research was supported by the Minerva Center 
for Nonlinear Physics of Complex systems.
}

\References

\endrefs


\begin{thebibliography}{99}
\bibitem{comp} Comparing to \cite{thomas} the main differences 
are: The $d{=}2$ diagram should be distinguished from the 
$d{=}1$ diagram since the former should include an intermediate
regime; The role of Heisenberg time for $d{=}3$ should be manifest 
in the ohmic regime; The localization axis should be 
appropriately placed.
\end{thebibliography}
\end{document}